\documentstyle[art12,epsf]{article}

\begin{document}

\title{Scaling above the upper critical dimension in Ising Models}
\author{Giorgio Parisi and Juan J. Ruiz-Lorenzo\\[0.5em]
{\small  Dipartimento di Fisica and Infn, Universit\`a di Roma}
   {\small {\em La Sapienza} }\\
{\small   \ \  P. A. Moro 2, 00185 Roma (Italy)}\\[0.3em]
{\small   \tt parisi@roma1.infn.it  ruiz@chimera.roma1.infn.it}\\[0.5em]}

\date{\today} 

\maketitle

\begin{abstract}

We rederive the finite size scaling formula for the
apparent critical temperature by using Mean Field Theory for the Ising
Model above the upper critical dimension. We have also performed
numerical simulations in five dimensions and our numerical data are in
a good agreement with the Mean Field theoretical predictions, in
particular, with the finite size exponent of the connected
susceptibility and with the value of the Binder cumulant.

\end{abstract} 

\thispagestyle{empty}

\newpage

\section{\protect\label{S_INT} Introduction}

In this letter we will study the scaling above the upper critical
 dimension for Ising Models (where Mean Field holds). This issue was
been controversial because the numerical simulations do not support
the Mean Field predictions \cite{BINDER}. 
In particular neither the scaling exponent
of the apparent critical temperature, nor the scaling exponents of the
connected susceptibility, nor the value of the Binder cumulant were
those of the Mean Field \cite{BREZINN,ZINN}. 
The first discrepancy has been solved using the
renormalization group \cite{BLOTE} and we will show that 
it can also  be explained using  Mean Field Theory. 
We have also made numerical simulations in five
dimensions and we see that our numerical values are in a good
agreement with the Mean Field predictions  and hence there is no
discrepancy.

The plan of this letter is the following. In the next section we will
made some remarks on the different limits  of the Binder cumulant. 
We will then derive the scaling formula for the  apparent critical
temperature using Mean Field Theory. 
Next  we show the numerical results
for the five dimensional Ising model and finally we will report the
conclusions.

\section{\protect\label{S_BIN} Limits of the Binder Cumulant}

The problem of the triviality of the field theory can be understood
studying the value of the renormalized coupling constant. We center
the following discussion using one component spin models.

If we
renormalize the theory at zero momenta it can be shown that the
renormalized constant, $g_R$, obeys:
\begin{equation}
g_R=\left( \frac{L}{\xi_\infty}\right)^d B ,
\end{equation}
where $\xi_\infty$ is the correlation length defined through the decay of
the correlation function, $d$ is the dimension,
and $B$, the Binder cumulant, defined as:
\begin{equation}
B=\frac{1}{2} \left (3-\frac{\langle M^4 \rangle}{\langle M^2 \rangle
^2} 
\right) ,
\end{equation}
where $M$ is the total magnetization. With this definition if the
distribution of $M$ is Gaussian $B$ vanishes.

From this point of view, triviality is equivalent to have a
renormalized coupling constant, $g_R$, which vanishes in the
thermodynamical limit at the critical point (where we define the field
theory).  But $g_R$ equal to zero implies that the Binder cumulant is
also zero. This corresponds to requiring that the limiting (infinite
volume) distribution of the total magnetization is Gaussian, as
happens, for instance, in the high temperature phase. In a non trivial
theory, the limiting probability distribution of the magnetization is not
Gaussian, and the degree of the non-gaussianity is measured  by the value
of the renormalized constant. In the low temperature phase the limiting
distribution is the sum of two Dirac deltas (one with positive
magnetization and the another with negative) and then the Binder
cumulant is one \cite{GIORGIO1}.

In this context is clear that we can use the value of $g_R$ to
classify the universality class and since $L/\xi_L$ is a
constant at $\beta_c$ we could also use the Binder cumulant for the
same purpose.

The problem is that we measure the Binder cumulant on a
lattice $L$ at temperature $1/\beta$, and so we obtain a function
$B(L,\beta)$. 
The finite size scaling behavior of the Binder cumulant is
\begin{equation}
B(L,\beta)=f(L^{1/\nu} (\beta-\beta_c)) .
\label{scaling}
\end{equation}
The scaling function has the following limits:
\begin{itemize}
\item{$\beta$ is fixed to the critical value $\beta_c$ and then  we 
take the infinite volume limit: 
\begin{equation}
B_1 \equiv \lim_{L\rightarrow \infty} \lim_{\beta\rightarrow \beta_c} 
B(L,\beta). 
\end{equation}
}

\item{We take the value of
the Binder cumulant at the apparent critical temperature,
denoted as $\beta_c(L)$ (defined, for example, as the maximum of the
connected susceptibility, obviously 
$\beta_c=\lim_{L\rightarrow \infty} \beta_c(L)$), and we consider the limit:
\begin{equation}
B_2\equiv\lim_{L\rightarrow \infty} B(L,\beta_c(L)) .
\end{equation} 
}
\item{First take $L\rightarrow \infty$ and then $\beta \rightarrow \beta_c$:
\begin{equation}
B_3 \equiv\lim_{\beta\rightarrow \beta_c} \lim_{L\rightarrow \infty} 
B(L,\beta) .
\end{equation}
}
\end{itemize}

 In general, these limits are all different:
\begin{equation}
B_1 \neq B_2 \neq B_3.
\end{equation}

Assuming that we have the following scaling for the apparent critical
point, $\beta_c(L)$
\begin{equation}
\beta_c(L)=\beta_c +a L^{1/\nu},
\end{equation}
where $a$ is dependent on the precise definition of $\beta_c(L)$ 
(for example the specific
heat or the susceptibility). Then we have:
$$
B_2=f(a).
$$
Obviously $B_2$ depends on the observable used to
define the apparent critical temperature.  This limit can be related to
the usual calculations of the Binder cumulant at $L/\xi_L={\rm const}$:
at the maximum of the connected susceptibility in a finite size we
have that the ratio between $\xi_L$ and $L$ is constant, independent
of the size.

We  remark that if we rewrite equation (\ref{scaling}) as
\begin{equation}
B(L,\beta)={\tilde f}(L/\xi_L(\beta)),
\label{scaling1}
\end{equation}
and  calculate the Binder cumulant for a set of $\beta$'s and $L$'s such
that $L/\xi_L(\beta)={\rm const}$, the value of this Binder cumulant will be
independent of the $\beta$'s and $L$'s in the scaling region (defined as the
region where the formula (\ref{scaling}), or equivalently
(\ref{scaling1}), holds).

For completeness:
\begin{equation}
B_1=f(0) \;\;,\;\;\; B_3=f(\infty).
\end{equation}
$B_1$ and $B_2$ are both size independent and observable independent: 
are the zero value and the infinite value, respectively, 
of a universal function.

In next section we will show how $B_2$ can be calculated using 
the connected susceptibility to define $\beta_c(L)$. 

\section{\protect\label{S_A} Above the critical dimension.}

We will follow the notation  of reference \cite{ZINN} in our
presentation. Above the critical
dimension we can study the system considering an effective action with
only one degree of freedom
\begin{equation}
S_{\rm eff}=L^d \left( \frac{1}{2} t \phi^2 +\frac{1}{4!} u \phi^4 \right),
\label{s_eff}
\end{equation}
where $t$ is the reduced temperature and we need to calculate:
\begin{equation}
m_{2p}= \frac{1}{\cal Z}\int {\rm d}\phi\; \phi^{2p} \exp(-S_{\rm eff}),
\end{equation}
where
\begin{equation}
{\cal Z}=\int {\rm d}\phi \exp(-S_{\rm eff}).
\end{equation}
In this notation the Binder cumulant is simply 
\begin{equation}
\label{binder_cumulant}
B=(3-m_4/m_2^2)/2.
\end{equation}
We perform the following change of variables:
\begin{equation}
\Phi=(u L^d)^{-1/4} \phi,
\end{equation}
and the action becomes
\begin{equation}
S_{\rm eff}= \frac{1}{2} x(t,L) \Phi^2 +\frac{1}{4!} \Phi^4 ,
\label{ACT}
\end{equation}
where $x(t,L)=t L^{d/2}u^{-1/2}$.

We remark that at this moment we are in a finite size and at a generic
temperature, different from the critical one.

If we calculate the value of $x(t,L)$ at the maximum of the connected
susceptibility ($<\phi^2>-<|\phi|>^2$) with the action of the
equation (\ref{ACT}) (in other words, we calculate the
apparent critical reduced temperature $t(L)$) 
 we obtain,  evaluating the integrals numerically:
\begin{equation}
x(t,L)_{\rm max}= -0.44015= t(L) L^{d/2} u^{-1/2}.
\end{equation}
and hence the scaling of the apparent reduced temperature is
\begin{equation}
t(L) \sim L^{-d/2},
\end{equation}
as was previously  found in
reference \cite{BLOTE} and numerically in \cite{BINDER}. The
subdominant term $L^{2-d}$ follows from the subsequent 
analysis of Brezin and Zinn-Justin \cite{BREZINN}. 
The analysis of
Brezin and Zinn-Justin was done with $t=0$ and they miss 
the dominant term $L^{-d/2}$ because it is proportional to $t$.

We can evaluate the Binder cumulant, eq. (\ref{binder_cumulant}),
 at this apparent critical
temperature with the action (\ref{s_eff})  obtaining
\begin{equation}
B_2=B(L,\beta_c(L))=0.742137.
\label{Binder}
\end{equation}
This value  is independent of the
dimension, size  and of the coupling $u$ but depends of the 
observable used (in this case the connected susceptibility). 
The generalization to
${\rm O}(N)$ models is straightforward.

\section{\protect\label{S_5} Numerical Results in $d=5$.}

We have performed numerical simulations in order to check the
discrepancy between previous quoted values \cite{BINDER} and the
theoretical prediction.

\begin{figure}[htbp]
\begin{center}
\leavevmode
\epsfysize=250pt
\epsffile{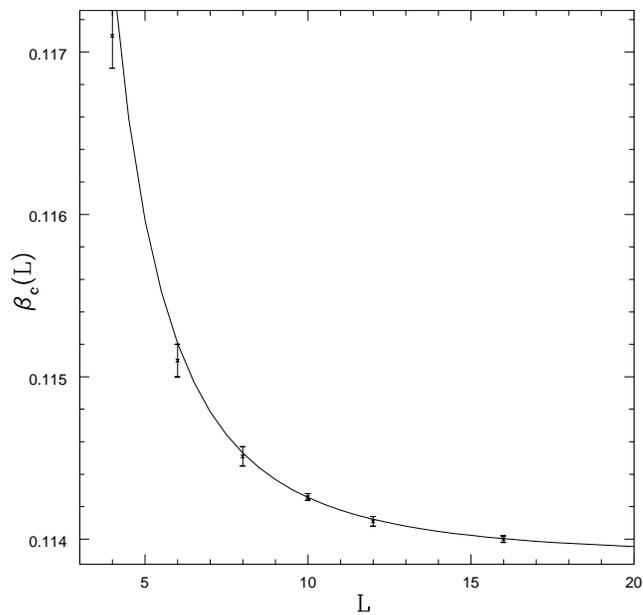}
\end{center}
\caption{Effective inverse temperature, $\beta_c(L)$, against $L$ for six 
different sizes.The continuous line is the global fit
supposing that the exponent is 2.5 (as described in the text). }
\end{figure}

To do this we have simulated  the five dimensional Ising Model 
using the Wolff single
cluster algorithm. We have simulated $L=4,6,8,10,12$ and 16 for a
total number of cluster sweeps of the order 6.5 million.  In the analysis
of the data we have used the spectral density method \cite{SDM} and
the jack-knife method to evaluate the statistical error.
\begin{table}
\begin{center}
\begin{tabular}{|c|c|c|} \hline
$L$ & $B(L,\beta_c)$ & $B(L,\beta_c(L))$\\ \hline \hline
4   &   0.485(5)  & 0.739(3)    \\ \hline
6   &   0.475(8)  & 0.744(4)    \\ \hline
8   &   0.450(2)  & 0.739(7)    \\ \hline    
10  &   0.425(5)  & 0.746(4)    \\ \hline
12  &   0.39(2)   & 0.73(1)    \\ \hline
16  &   0.40(2)   & 0.74(1)    \\ \hline
\end{tabular}
\caption{$B(L,\beta_c)$ and $B(L,\beta_c(L))$ for different sizes. 
The Mean Field theoretical value  is 0.40578 for the first Binder
cumulant and 0.742 for the second one.}
\end{center}
\end{table}

The shift of the critical temperature, calculated as the maximum of
the connected susceptibility, follows (using all the sizes in the fit)
\begin{equation}
\beta_c(L)=0.11387(4) +0.0798(20) L^{-2.32(16)},
\end{equation}
with a $\chi^2/{\rm d.o.f} \approx 0.2 $ and we write the errors inside of
the parenthesis.
Hence we obtain a shift compatible with the theoretical prediction
$5/2$.  This value was also
reported in \cite{BINDER}. We have repeated the global fit with $L\ge
6$ the numbers are very similar ($\beta_c=0.11386(6)$) but the errors are
large.

\begin{figure}[htbp]
\begin{center}
\leavevmode
\epsfysize=250pt
\epsffile{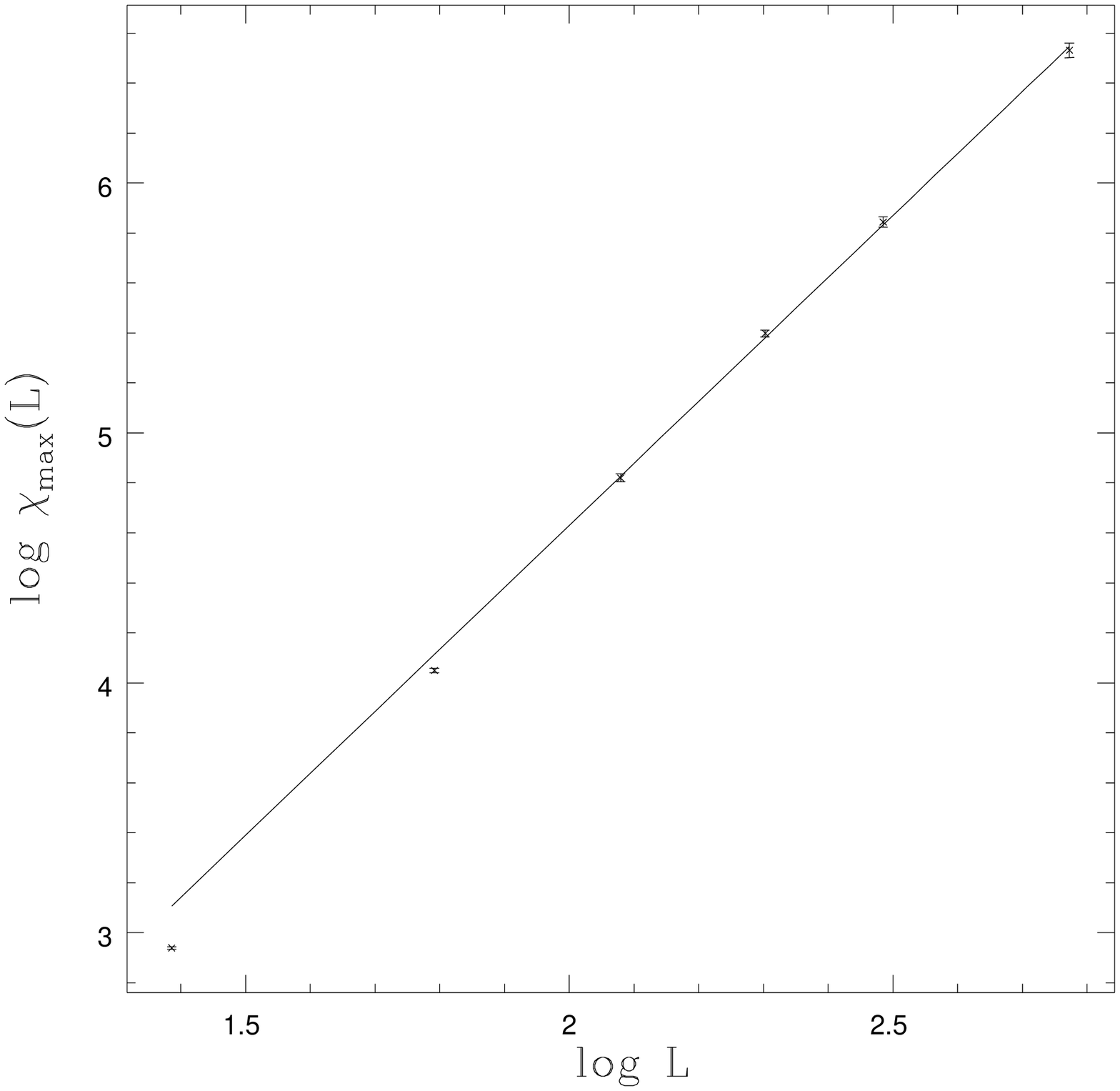}
\end{center}
\caption{Scaling of the connected susceptibility for six different
sizes. The continuous line is the fit using only the points $L \ge 8$,
as described in the text.}
\end{figure}

Fixing the scaling exponent to the
theoretical value of eq. (2.5) and using $L\ge 8$ , we obtain:
\begin{equation}
\beta_c(L)=0.11388(3)+0.12(1) L^{-2.5}
\label{fit1}
\end{equation}
with a good $\chi^2/{\rm d.o.f.}\approx 0.15$.
In Fig. 1 we plot the numerical values of $\beta_c(L)$ and plot,
signed by a continuous line, the fit (\ref{fit1}). We remark that the 
estimation of Rickwardt et al \cite{BINDER} was $\beta_c=0.113929(45)$
that is a 1.5 standard deviations of our value.

\begin{figure}[htbp]
\begin{center}
\leavevmode
\epsfysize=250pt
\epsffile{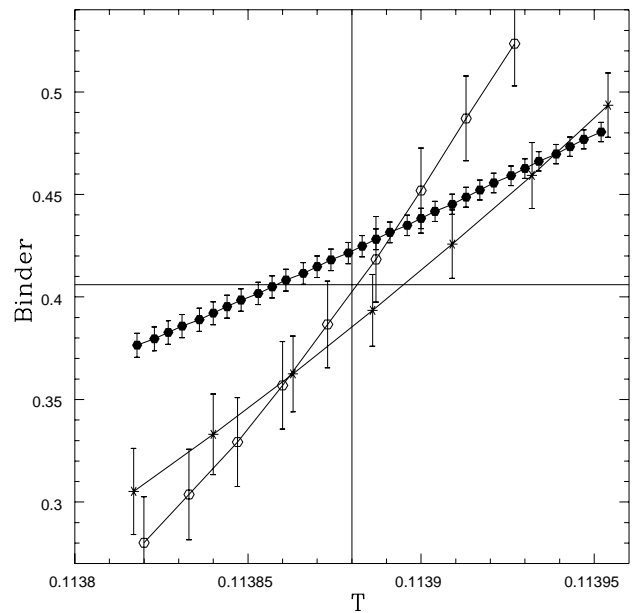}
\end{center}
\caption{ We plot the curves of the Binder cumulant for $L=10$
($\bullet$), $L=12$ ($\ast$)  and
$L=16$ ($\circ$). The  horizontal line is the Mean Field prediction
(B=0.406). The vertical line is our estimation of the inverse critical
temperature.}
\end{figure}

Using $\beta_c=0.11388$ we can monitor $B(L,\beta_c(\infty))$ that we
plot in Table 1. For the large lattices the value is compatible
with that reported by Brezin and Zinn-Justin in reference
\cite{BREZINN}, $B=0.40578$. Also we have written in this Table the values
of $B(L,\beta_c(L)$ and the agreement is again very good with our
previous calculated value (\ref{Binder}).

Analyzing the maximum value of the connected susceptibility we found 
the following power law using all the lattices in the fit:
\begin{equation}
\chi_{\rm max} \sim L^{2.687(6)},
\end{equation} 
but the  fit is very poor ($\chi^2/{\rm d.o.f.}=7$). However  
fitting only the lattices with  $L \ge 8$
we found   
\begin{equation}
\chi_{\rm max} \sim L^{2.48(4)},
\label{fit2}
\end{equation}
with $\chi^2/{\rm d.o.f.} \approx 1.95/2$, 
and so the agreement is very good with the
theoretical prediction $5/2$.  We plot the numerical data with the fit
(\ref{fit2}) in Fig. 2.

We finally report in Fig. 3 the different curves of the Binder
cumulant near the critical temperature. Using the critical temperature
of reference \cite{BINDER} we have found, using our $L=16$ data, a
value of the Binder cumulant $B=0.52(2)$ that is a two standard
deviations of $B=0.479(25)$, the value reported by this group
\cite{BINDER}.

\section{\protect\label{S_CONCLU} Conclusions.}

We have derived using Mean Field the finite size shift of the apparent
critical temperature for the Ising model above the upper critical
dimension. We also have shown that the numerical data for the five
dimensional Ising model are compatible with the Mean Field
predictions, in particular, the value of the Binder cumulant at
criticality, the finite size exponent of the connected susceptibility
and the shift of the apparent critical temperature agree very well
with Mean Field Theory.

\section*{Acknowledgments}

We acknowledge useful discussions with D. J. Lancaster,
L. A. Fern\'andez and E. Marinari.

\end{document}